\documentclass[aps,prl,twocolumn,superscriptaddress,showpacs,english]{revtex4-1}

\usepackage{float}
\usepackage{caption}
\usepackage{subcaption}
\usepackage[T1]{fontenc}
\usepackage[latin9]{inputenc}
\usepackage{babel}
\usepackage{amsmath}
\usepackage{amssymb}
\usepackage{wasysym}
\usepackage{graphicx}
\usepackage{multirow}
\usepackage{gensymb}
\usepackage[dvipsnames]{xcolor}
\usepackage{pifont}
\usepackage{microtype}
\usepackage{xcolor}
\usepackage{soul}
\usepackage[version=3]{mhchem} 
\usepackage{chemformula}[2014/04/08] 
\usepackage{booktabs}    
\usepackage{dcolumn}     
\usepackage{bm}          
\usepackage[linktocpage=true,
  colorlinks=true, 
  pdfborder={0 0 0},
  linkcolor=blue,
  citecolor=red,
  filecolor=yellow,
  urlcolor=blue,
  bookmarks,
  pdfauthor={},
]{hyperref}

\begin{document}

\title{Detecting non-local effects in the electronic structure of a simple covalent system with machine learning methods}

\author{Behnam Parsaeifard}
\author{Jonas A. Finkler}
\author{Stefan Goedecker}

\affiliation{Department\ of\ Physics,\ University\ of\ Basel,\ Klingelbergstrasse\ 82,\ CH-4056\ Basel,\ Switzerland}
\affiliation{National Center for Computational Design and Discovery of Novel Materials (MARVEL), Switzerland}  


\begin{abstract}
Using methods borrowed from machine learning we detect in a fully algorithmic way 
long range effects on local physical properties in a simple covalent system of carbon atoms.
The fact that these long range effects exist for many configurations implies 
that atomistic simulation methods, such as force fields or modern machine learning schemes, that are based on locality assumptions, are limited in accuracy.
We show that the basic driving mechanism for the long range effects is charge transfer. If the
charge transfer is known, locality can be recovered for certain quantities such as the band structure energy.
\end{abstract} 

\maketitle




Most approximate chemical simulation schemes are based on a locality assumption. A local property, such as a local charge distribution, 
an atomic spin polarization or atomic energy as well as bond lengths  are assumed to depend only on a nearby local environment 
but not features far away. The locality assumption is very well satisfied in many covalently bonded systems.
As an example let us consider the total energy of the alkanes polymers, $C_n H_{2n+2}$. 
Each $CH_2$ monomer is, energetically 
virtually an independent unit. As one
adds an additional $CH_2$ monomer, the energy increases by an
amount that is nearly independent of the chain length. 
Insertion of a $CH_2$ monomer into the smallest
chain, $C_2 H_6$, gives already an energy gain that agrees to within
$10^{-4}$ Ha with the asymptotic value of the insertion energy for very long chains~\cite{ONrev}. 
This shows that the electrons
belonging to this inserted sub-unit no longer "see" the
end of the chain. This locality principle has therefore been dubbed 
"nearsightedness"  by Walter Kohn~\cite{Kohn1996} and he claimed it to be valid nearly universally. 
In this study we will consider pure carbon systems and show that even in such a simple covalent system 
non-local effects play an important role. 

All the standard force fields for this material~\cite{marks} such as 
EDIP~\cite{PhysRevB.63.035401}, Tersoff~\cite{PhysRevB.37.6991}, Brenner~\cite{PhysRevB.42.9458} or 
recent versions of bond order potentials~\cite{los2005improved} are also based on this locality assumption.  Modern machine learning schemes~\cite{Deringer,rupp,burke}, are based on this locality assumption as well.  The energy is given in these schemes 
as a sum over atomic energies which depend only on a short range environment. Long range electrostatic energies are 
sometimes still added~\cite{behler2015constructing} but the atomic charges giving rise to these interactions depend again 
only on a local environment whereas in reality they are strongly influenced by non-local effects. 

To demonstrate the existence of non-local effects, 
one has to show that local physical properties are different for short range environments that are virtually identical. 
Environment descriptors, also called atomic fingerprints,  that quantify the similarity of chemical environments have 
recently been developed in the 
context of machine learning and for analysing big structural data banks~\cite{behler2011atom,bartok2013representing,faber2018alchemical,sadeghi2013metrics}.
We will use in this study the fingerprints based on eigenvalues of an atom centered overlap matrix~\cite{zhu2016fingerprint} 
since these descriptors have demonstrated a high reliability in detecting differences in the local environment~\cite{parsaeifard2020assessment}.  
We use a cutoff radius of $6$~\AA \ and s and p type orbitals for the overlap matrix.
Denoting a fingerprint vector describing the environment of two atoms $\alpha$ and $\beta$ 
by $\bm f_{\alpha}$ and $\bm f_{\beta}$, we obtain a measure 
of the similarity by calculating the fingerprint distance as the euclidean norm $| \bm{f}_{\alpha} -  \bm{f}_{\beta}|$. 
Small values indicate that the environments are similar. 
In this work we will correlate fingerprint distances with differences of localized physical properties of the system such as
atomic charge densities, atomic energies, and atom-projected densities of states.
These changes in 
the charge densities will finally also modify bond lengths of our systems in a non-local way.

To split up global quantities into atomic quantities we use the following 
partitioning of the unity $W_\alpha(\bm r)$:
\begin{equation}
    W_\alpha(\bm r) = \frac{e^{-(\frac{\bm r -\bm R_\alpha}{\sigma})^2}}{\sum_{\beta}^{N_{at}}{e^{-(\frac{\bm r -\bm R_\beta}{\sigma})^2}} }
\end{equation}
where $N_{at}$ is the number of atoms in the system and $R_\alpha$ denotes the Cartesian coordinates 
of atom $\alpha$. $\sigma$ is some smearing parameter which we take to be equal to the covalent radius of atom $\alpha$. 
The function $W_\alpha(\bm r)$ has large values around atom $\alpha$ and as we move further away from atom $\alpha$ it becomes 
very small and it has the property $\sum_\alpha W_\alpha(\bm r)=1$. It can be considered as some kind of smooth Voronoi decomposition of space since it will 
give the Voronoi decomposition in the limit of small $\sigma$. 
Let us also still point out the trivial but important point that this smooth Voronoi decomposition depends only on the nearest neighbor positions. 
So if the local environment is not changed the Voronoi volume will not be modified either. 
Hence, if some quantity that is derived from this partitioning exhibits non-local effects 
it can not be due to some change in the shape of the smooth Voronoi volume but must be due to a change in the physical quantities. 
The physical quantities that will be examined
are the wavefunctions and their Kohn-Sham eigenvalues.
 
As a first quantity we define atomic charges $\rho_{\alpha}$
\begin{equation} \label{eq:rho}
  \rho_{\alpha}=\int{d\bm r  \sum_i {n_F(\epsilon_i) |\phi_i(\bm r)|^2 W_\alpha(\bm r)}   }   
\end{equation}
where $\epsilon_i$ and $\phi_i$ are eigenvalues and eigenfunctions of the Hamiltonian of the system and are obtained by 
solving the Schroedinger equation for the system within Density Functional Theory (DFT) as implemented in BigDFT~\cite{bigdft,willand} using the PBE functional~\cite{PBE}. 
$n_F(\epsilon)$ is the occupation number of the state with energy $\epsilon$ at an electronic temperature $k_B T$ of $10^{-5}$  Ha. 
Since, as pointed out above, the Voronoi volume will not be influenced by non-local effects, this quantity is a direct measure 
of the change in the charge density around the central atom. This is in contrast to some other charge decomposition schemes 
such as Bader~\cite{bader} 
or Mullikem~\cite{mulliken}, where the volume associated to an atom is not determined by the geometry of the local environment but by 
the charge density or the wavefunction. 

\begin{figure*}[tpbh]
    \centering
    \includegraphics[width=\textwidth]{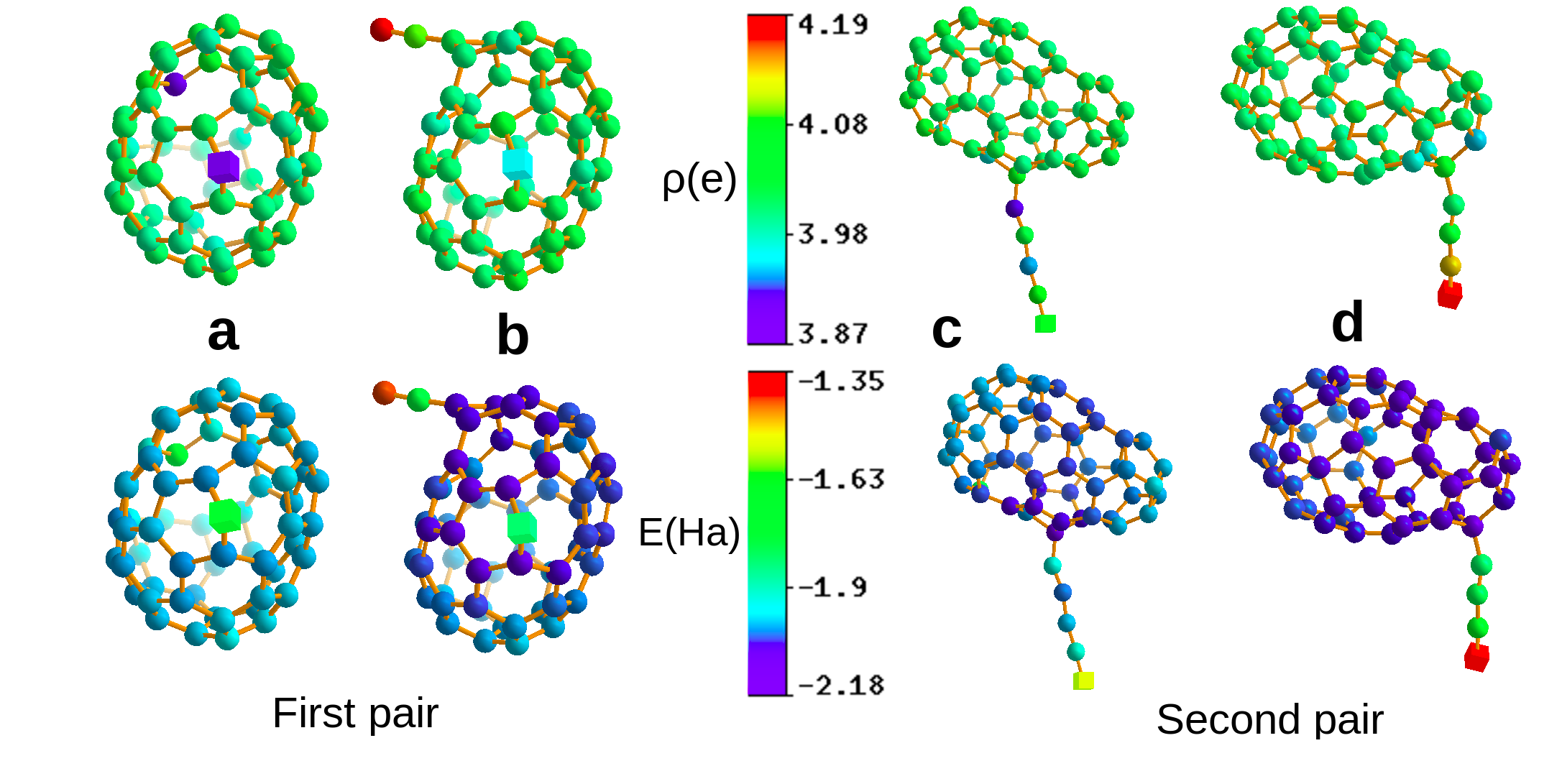}
    \caption{Two pairs of distinct structures, where we can find central atoms (shown as cubes) that 
    are in virtually identical short range chemical environments.
    with fingerprint distances of $\Delta FP_{OM} \approx 10^{-2}$. 
    Due to long range effects the atomic charges and atomic energies are however quite different. The atoms in the first and second row are colored according to their atomic charges 
    and energies respectively. $\Delta E=0.11$ Ha and $\Delta \rho=0.08$ electrons for the pair in the left 
    column and $\Delta E=0.21$ Ha and $\Delta \rho=0.12$ electrons for the pair in the right column.  }
    \label{fig:two_examples}
\end{figure*}

As a second quantity we define atomic energies $E_\alpha$. Since the decomposition of the total energy is highly ambiguous~\cite{NAKAI}, we
perform this decomposition only for the band structure energy  which can again be assigned in a unique way to the smooth Voronoi volumes by partitioning the energy density
\begin{equation}\label{eq:ebs}
    E_\alpha=\int{d\bm r  \sum_i { n_F(\epsilon_i) \epsilon_i  |\phi_i(r)|^2 W_\alpha(\bm r)}   }  
\end{equation}
Since $W_\alpha(\bm r)$ is a partitioning of the unity 
the sum over all the atomic energies gives the band structure energy, i.e. $\sum_{\alpha}^{N_{at}}{E_\alpha} =\sum_i {\epsilon_i}$.
As is well known~\cite{harris} the band structure energy term, $ \sum_i {\epsilon_i}$,  is the most important term 
to describe variations in the total energy.
As shown in Fig.~\ref{fig:two_examples} these atomic energies agree well with our basic chemical intuition of 
which environment will give rise to low or high atomic energies.  The atoms at the end of the chains have 
for instance the highest energies whereas the atoms 
of the cage have lower energies. For these atoms the energy is however also larger for atoms in a defective cage region


\begin{figure*}
    \centering
    \includegraphics[width=\textwidth]{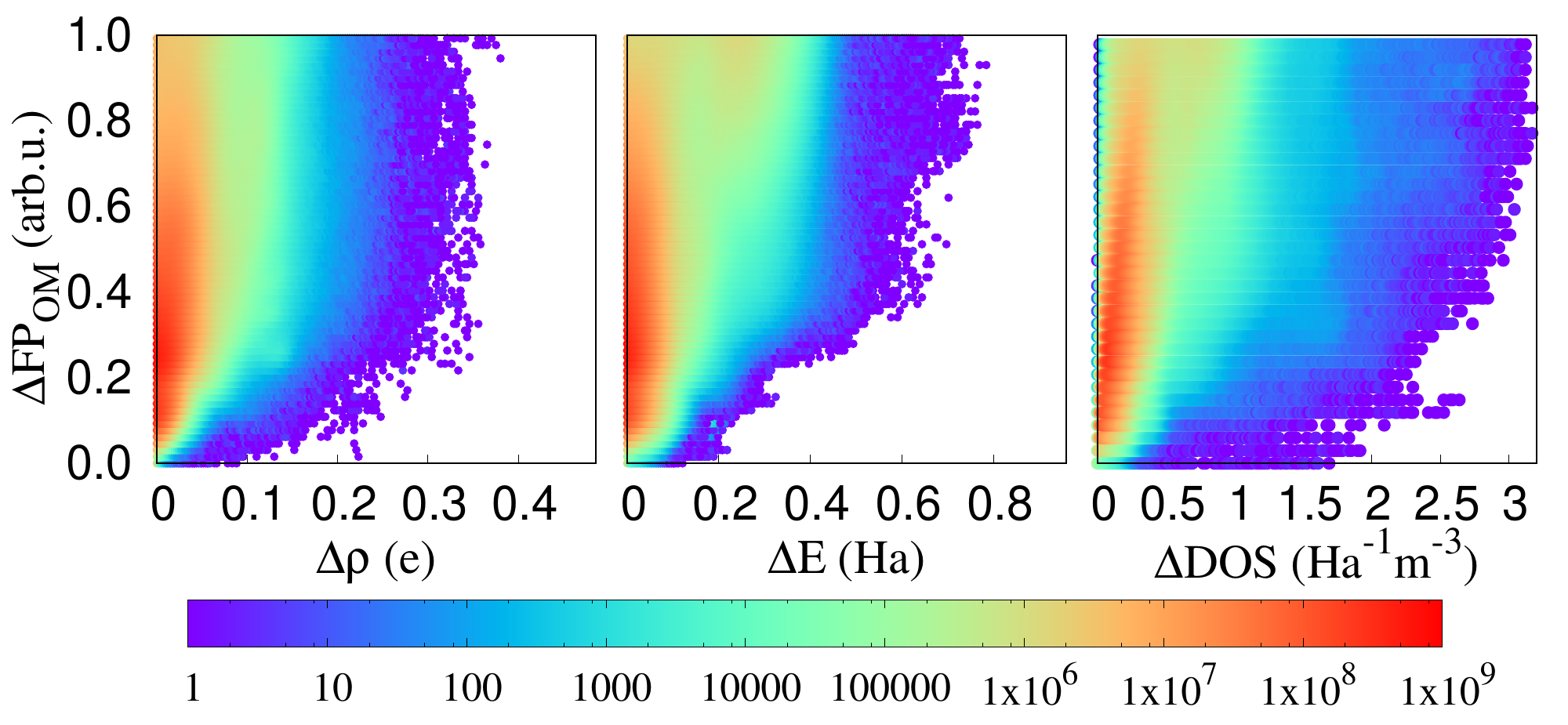}
    \caption{The correlation plot between OM fingerprint distances and differences in atomic charge (Eq. \ref{eq:rho}), atomic energy (Eq. \ref{eq:ebs}), and the atom projected DOS (Eq. \ref{eq:projected_dos}).}
    \label{fig:corr_nonlocal}
\end{figure*}
\begin{figure}
    \centering
    \includegraphics[width=\columnwidth]{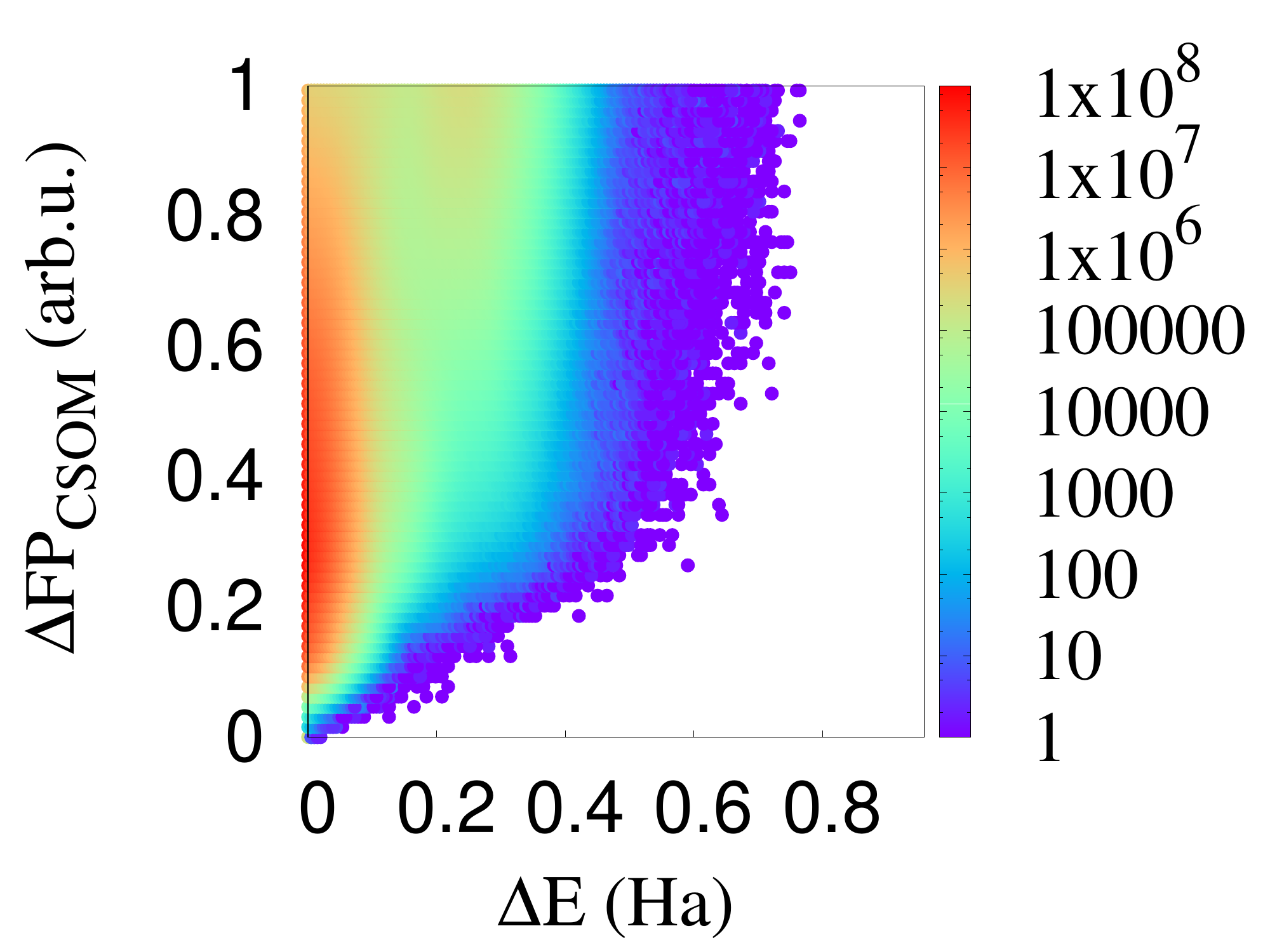}
    \caption{The correlation between distances calculated with the charge-sensitive OM fingerprint (CSOM) and the atomic energy differences.}
    \label{fig:corr_local}
\end{figure}

As a third quantity we study the atom projected density of states.
The density of states for the system is $D(\epsilon)=\sum_i {\delta(\epsilon-\epsilon_i)}$.
We define the atom-projected density of states for atom $\alpha$ to be:
\begin{equation} \label{eq:projected_dos}
    D_\alpha(\epsilon)=\int d\bm{r} \sum_i { \delta(\epsilon-\epsilon_i) n_F(\epsilon_i) |\phi_i(\bm r)|^2  W_\alpha(\bm r)}
\end{equation}
With the property $\sum_\alpha D_\alpha(\epsilon)=\sum_i \delta(\epsilon-\epsilon_i)=D(\epsilon)$.
We replace $\frac{1}{\sqrt{2\pi \sigma^2}} \exp \left(\frac{-\left(\epsilon-\epsilon_i\right)^2}{2\sigma^2}\right)$ for $\delta\left(\epsilon-\epsilon_i\right)$ in Eq.~\ref{eq:projected_dos} where $\sigma$ is a smearing parameter whose value is 0.05 Ha. We define the difference between the atom-projected density of states of two atoms $\alpha$ in structure $p$ and $\beta$ in structure $q$ to be: 
\begin{equation}
  \Delta DOS_{\alpha(p),\beta(q)}=\sqrt{\int d\epsilon  \left(D_{\alpha(p)}\left(\epsilon\right)-D_{\beta(q)}(\epsilon)\right)^2}
\end{equation}
This quantity can be calculated analytically for the  numerically obtained $\epsilon_i$'s. 

By a combination of Minima Hopping~\cite{minhop} and MD coupled to DFTB ~\cite{aradi2007dftb+}, 
we have generated a large number of clusters with 60 carbon atoms.
This data base of 3000  $C_{60}$ configurations  containing a wide range of structural motifs including 
chains, graphitic sheets  and cages. In this way 180000 environments were created.
By analysing the correlation between the fingerprint distances and the physical 
observables, we will show that it is possible to detect in a fully automatic way  non-local effects in our structures. 
So our search for non-local effects is much 
more comprehensive than it would be possible with a search based on human intuition. 


In Fig.~\ref{fig:corr_nonlocal} we plot differences of three local physical properties, 
namely atomic charges, atomic energies and the atom-projected density of states, against 
fingerprint distances. In all these cases it may happen that the same value of a physical property is 
observed for different environments. Energies might for instance be degenerate. However, if these localized physical
properties differ for  identical or nearly identical environments, localized physical properties are 
influenced by long range effects. Such cases correspond to points on or very close to the x axis in our correlation plots and we 
see that indeed plenty of such points exist. As shown in detail in the supplementary information (SI), the existence of these points is independent of the fingerprint used as long as the fingerprint has a high structural resolution. Hence, long range effects clearly  exist in this covalent system.


Having established the existence of long range effects on local physical properties in a purely algorithmic way, it is interesting 
to see whether they can also be explained by traditional physical arguments. 
A structure that is
strongly affected by non-local effects is the structure shown in panel \textbf{d} of Fig.~\ref{fig:two_examples}.  It consists of a cage 
of 56 carbon atoms and a 4 carbon atom chain attached to it.
If one calculates the Kohn Sham eigenvalues of the two isolated fragments, 
i.e the 4 atom chain and the 56 atom cage one finds that the LUMO level of the chain is lower than the HOMO level of the cage.
Hence, in a simple one particle picture one electron would be transferred from the cage to the chain. In a DFT calculation such a
charge transfer is always reduced by the electron-electron repulsion and based on our analysis of the atomic charges we find indeed 
only a charge transfer of about 0.34 electrons in that case. 
We were able to find analogous explanations for several other cases that 
we inspected in more detail, but not for all of them. 

It is for instance probably not possible to predict by basic chemical reasoning the variation on the atomic charge on the central atoms in the pair of structures shown in panel \textbf{c} and \textbf{d} of Fig.~\ref{fig:two_examples}. 
Both central atoms are the outermost ones in a chain attached to a cage 
and the cage structures look quite similar.
Hard to explain by traditional arguments are also the differences in the charge of the two central atoms shown in panel \textbf{a} and \textbf{b} of Fig~\ref{fig:two_examples} where again the near environments are almost identical and only the structure of the cage differs slightly. 
So this shows that our purely algorithmic approach is actually more powerful in 
detecting non-local effects than the traditional approaches.

Varying atomic charges are supposed to lead to variations in the bond length and this is indeed the case for this system.
The bond lengths of the 4 atom chain differ depending on whether the chain is isolated or attached to the cage. 
The bond lengths of the PBE relaxed free chain of 4 carbon atoms are 1.293, 1.313 (middle bond), and 1.293.
For the attached chain (chain being attached to the cage in Fig.~\ref{fig:two_examples} \textbf{d}) the PBE bond lengths change to 1.243, 1.327 (middle bond), and 1.280 (the bond at the free end of the chain). 
So, the bond length at the free end of the chain becomes significantly shorter due to the transferred charge. In addition the electronic ground state of 
the free chain is also spin polarized. So, long range effect modify both the bond lengths and the spin moments. 

Having established the ubiquitous existence of non-local effects in a standard covalent material, one has to question whether  the near sightedness postulated by Walter Kohn holds.
Actually in the publication where this notion of near-sightedness was introduced there is a caveat,  namely  that it is only valid 
if the chemical potential is constant.  Since  charge transfer is driven by a varying chemical potential this principle is therefore 
not directly applicable in real systems where, as shown in this study, such a charge transfer is quite common.
Because of its central importance in the calculation of the total energy, we will in the following 
concentrate on the atomic band structure energy and show that locality can be restored for this quantity  
if one includes not only information about the structure in a limited environment but also about the atomic charges.
For this purpose we modify 
our fingerprint such that it also  depends on the atomic charges within the sphere with our 
chosen cutoff radius as detailed in the SI.
In this way the resulting fingerprint still has a strong sensitivity to the geometrical structure 
but in addition a weak sensitivity to the charges.
As can be seen from Fig.~\ref{fig:corr_local} all the points which were in Fig.~\ref{fig:corr_nonlocal}
close to the x-axis are now moved upward. Hence there are no more additional long range effects.
This means that the charge transfer is the basic long-range effect. Once this charge transfer is known the total energy can be obtained from purely local information. 
This follows from the fact that the total energy can be obtained from electrostatic and exchange 
correlation terms that depend only on the charge density plus the band structure energy.
This finding has important consequences for machine learning schemes.
Charge transfer is not possible in most of the these schemes.
Hence they will necessarily be limited in accuracy. For instance the environment descriptors of the atoms at the end of the chains in Fig.~\ref{fig:two_examples}  
would have in all standard machine learning schemes a cutoff range which is shorter than the length of the chain. Hence the standard descriptor can not see whether the chain is free standing or attached to the cage. Some long range fingerprints that might cope with this deficiency have however also been proposed recently~\cite{mallat,grisafi2019incorporating}. 
Non-local charge transfer effects in combination with standard short range fingerprints can however be described by the CENT scheme~\cite{ghasemi2015interatomic,amsler2020flame}  where a machine learning scheme is combined with a charge equilibration scheme. 
Consequently a scheme of this type has to be an integral part of any machine learning scheme that 
strives to obtain very high accuracy also for systems where long range effects can not be neglected.

This research was performed within the NCCR MARVEL, funded by the Swiss National Science Foundation. The calculations were performed on the computational resources of the Swiss National Supercomputer (CSCS) under project s963 and on the Scicore computing center of the University of Basel.

\bibliography{main}

\end{document}